%% file: NSSDFIP-arXiv.tex
\documentclass[letter,twocolumn]{jpsj2}
\usepackage{bm}
\usepackage{graphicx}
\usepackage{subfigure}

\title{Normal State Spin Dynamics of Five-band Model for Iron-pnictides}
\author{Toshikaze \textsc{Kariyado} and Masao \textsc{Ogata}}
\inst{Department of Physics, Graduate School of Science, 
University of Tokyo, 7-3-1, Bunkyo-ku, Tokyo 113-0033\\
JST, TRIP, 5 Sanbancho, Chiyoda, Tokyo 102-0075}

\abst{ Normal state spin dynamics of the recently discovered 
 iron-pnictide superconductors is discussed by calculating spin
 structure factor $S(q,\omega)$ in an itinerant five-band model within RPA
 approximation. Due to the characteristic Fermi surface structure of
 iron-pnictide,
 column like response is found at $(\pi, 0)$ in extended Brillouin
 zone in the undoped case, which is consistent with the recent neutron
 scattering experiment. This indicates that the localized spin model is
 not necessary to explain the spin dynamics of this system. Furthermore,
 we show that 
 the temperature dependence of inelastic neutron scattering
 intensity can be well reproduced in the itinerant model.
 We also study NMR $1/T_1T$ in the same footing calculation and show
 that the itinerant model can capture the magnetic property
 of iron-pnictide superconductors.
 }

\kword{iron-pnictide, multi-orbital Hubbard model, 
inelastic neutron scattering, NMR}

\begin{document}
\maketitle

%introduction
Recently discovered\cite{Kamihara:2008} iron-pnictide (or
transition-metal pnictide) 
superconductors have attracted much attention and have been studied
intensively
since they show superconductivity at rather high temperatures up to 55K
\cite{Ren:2008b,Kito:2008} at present. Novelty of
this series of compounds is not limited to their
high Tc: entangled band structure, coexistence of hole and electron
Fermi surfaces\cite{Singh:2008,Kuroki:2008}, origin of the
magnetism\cite{Ishibashi:2008,Hsieh:2008},
interplay between magnetism and
superconductivity\cite{Drew:2008,Yanagi:2008b}, role
of lattice structure\cite{Lee:2008a,Vildosola:2008}, nodal or nodeless
behavior in the
superconducting
phase\cite{Nakai:2008b,Checkelsky:2008,Hashimoto:2008,Ding:2008c},
robustness
against impurities in the conducting
layer\cite{Sefat:2008b,Kawabata:2008,Senga:2008} and
so on. All
of these features enrich the physics of this compound. In this paper, we
pay attention to magnetic properties of this series of compounds,
especially the normal state spin dynamics.

% NMR and INS
For the study of spin dynamics, NMR and inelastic neutron scattering
experiments are powerful tools. There already exist systematic works of
NMR\cite{Nakai:2008b,Ning:2008a}, and some results of inelastic neutron
scattering on iron-pnictides
\cite{Ewings:2008,McQueeney:2008,
Zhao:2008e,Ishikado:2008,Matan:2008}. 
The normal state behavior of $1/T_1T$ obtained in the
NMR shows a clear tendency against electron doping. In the
low doping region, which is near the magnetically ordered phase, one can
see enhancement of $1/T_1T$ with decreasing temperature, while in the
high doping region, there is no enhancement of $1/T_1T$ and rather, it
decreases with decreasing temperature. On the other hand, in the
inelastic neutron scattering experiments, at least two common features
have been observed: one is column-like response at the antiferromagnetic
wavenumber and another is gapped spin excitation spectra.

%motivation and direction
Analysis of these results in the inelastic neutron scattering is often
carried out with a Heisenberg-type spin model\cite{Unrig:2008,Yao:2008a},
although this system shows
itinerancy. Sometimes this system is regarded as a bad metal, but it is
basically metallic in transport properties\cite{Liu:2008} (at most
semiconducting). In
addition, ARPES measurements show that the band structure coincides with
those obtained in the first principle calculation with a mass
renormalization factor of 2-4\cite{Malaeb:2008,Ding:2008}. This means
that, although electron correlation plays important role in the
iron-pnictides, it is not so strong as in the high Tc cuprates, whose
mother compounds are well described by a localized-spin Heisenberg
model. Thus, in this paper we analyze the inelastic neutron
measurements based on a purely itinerant model. We also discuss NMR
results of $1/T_1T$ briefly to be more concrete.

\begin{figure}[t]
\begin{center}
 \subfigure[extended]{\includegraphics[width=50pt]{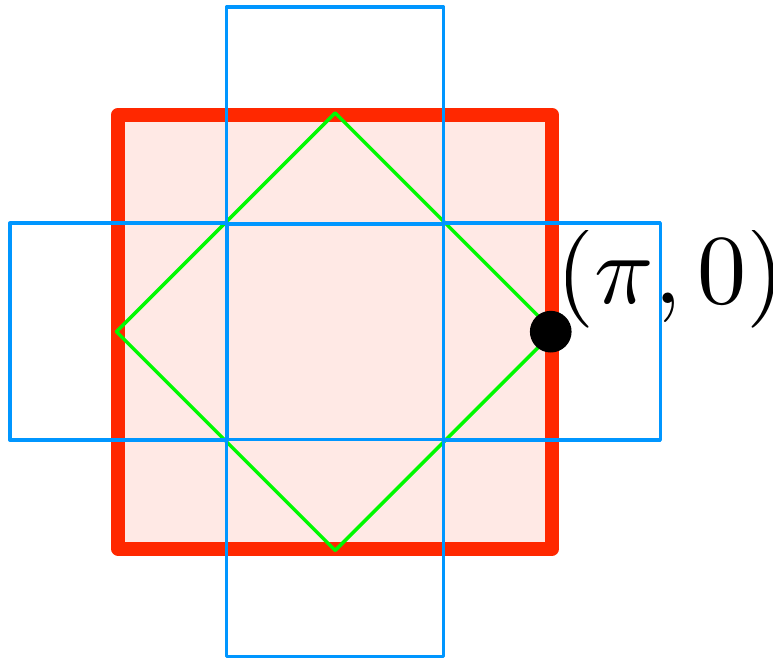}}
 \subfigure[original]{\includegraphics[width=50pt]{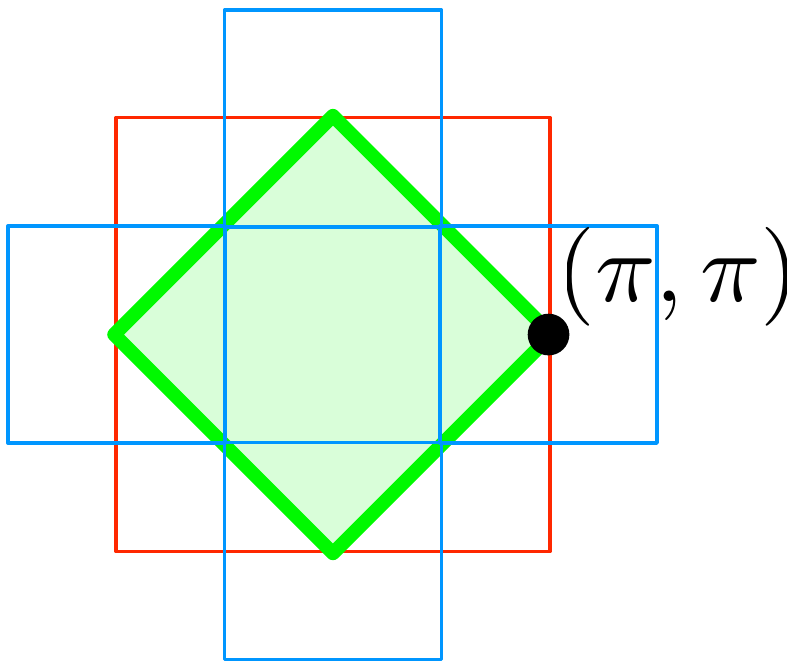}}
 \subfigure[magnetic]{\includegraphics[width=50pt]{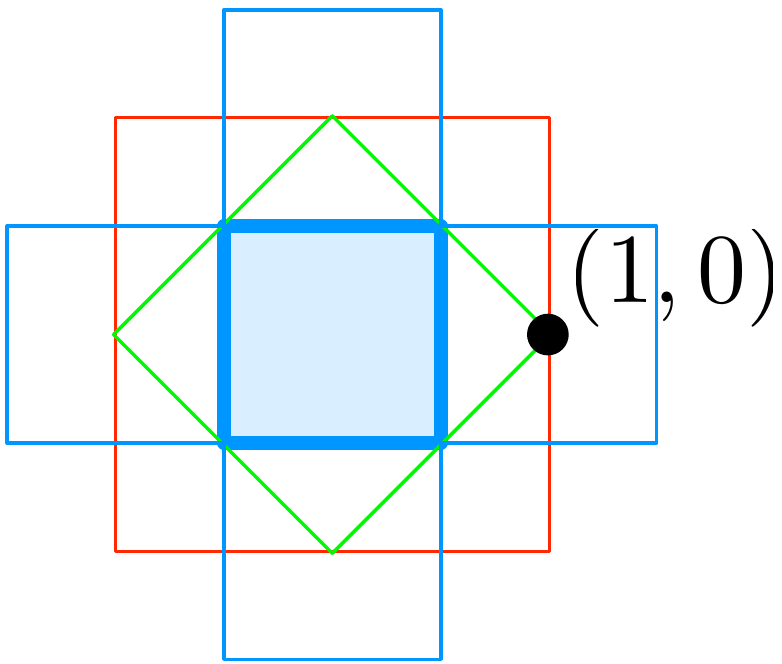}}\\
 \includegraphics[width=200pt]{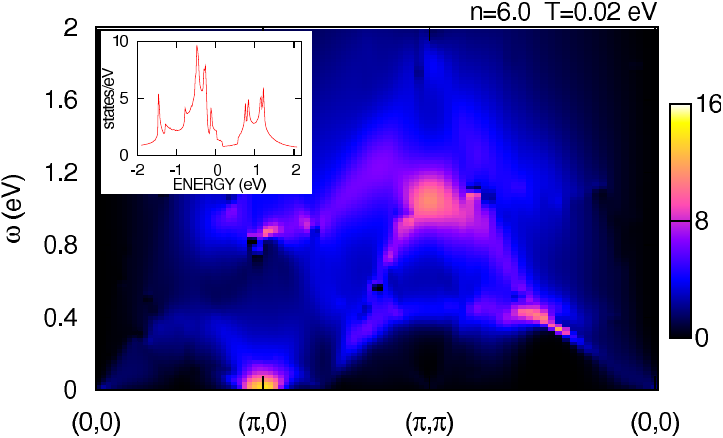}
 \caption{(Color online) Overall view of $S(\bm{q},\omega)$ along the
 high symmetry
 lines in the Brillouin zone at temperature $T=0.02$eV. Inset shows DOS
 of the present model. The upper panels
 denote the Brillouin zone of the present model
 containing one Fe per unit cell (a), original Brillouin zone having two
 Fe per unit cell (b), and that in the magnetic phase (c).}
 \label{fig1}
\end{center}
\end{figure}

%model and method
The model we use is the five-band Hubbard model proposed by Kuroki 
{\it et al.}\cite{Kuroki:2008} which is down-folded from the first principle
calculation. Five bands in this model mainly comes from five Fe-3d
orbitals and hopping parameters are kept up to fifth nearest neighbor. 
The dispersion relation and the structure of the density of states
are in good agreement with the first principle calculation. Although this
model is two-dimensional, this simplification is not so harmful
since the Fermi surfaces of this system are basically cylindrical. We must
also note that this model contains one Fe atom per unit cell and
Brillouin zone is doubled compared with the original one. (See
upper panel of Fig.~\ref{fig1}.) Actually, this model is constructed
from LaFeAsO system, but we believe that the properties of 
FeAs layer is qualitatively same for other types of iron-pnictide. The
interaction is limited to the onsite interaction and treated within RPA
approximation. We tried several sets of interaction parameters keeping
the constraints of
$U=U'+J+J'$ and $J=J'$, but the results shown in the following are
those with $U=1.1$, $U'=0.9$, $J=0.1$ and $J'=0.1$. Here, the unit of
the energy is electron volt (eV). We treat
doping as a rigid band shift, i.e., we neglect possible structural change
due to the change of the charge balance between the conduction layers and
blocking layers. Note that $n=6.0$ corresponds to the stoichiometric
sample in this model. We
divide Brillouin zone into 64$\times$64 meshes and 1024 Matsubara
frequencies are used. Pad\'{e} approximation is used to transform the
results on Matsubara frequencies to those on real frequency.

%S(q,w) definition
In order to analyze the inelastic neutron scattering experiments, we calculate 
spin structure factor $S(\bm{q},\omega)$, which is defined as Fourier
transform of a spin correlation function,
\begin{equation}
 S(\bm{q},\omega)=\frac{1}{N}\sum_{\bm{r}}\mathrm{e}^{-i\bm{q}\cdot \bm{r}} 
  \int^{\infty}_{-\infty}\mathrm{d}t\mathrm{e}^{-i\omega t}
  \langle\hat{S}_{\bm{0}}\cdot\hat{S}_{\bm{r}}(t)\rangle.
\end{equation}
To be precise, we must include a form factor, often denoted as
$F(\bm{q})$, in the above definition for the comparison with
experimentally obtained scattering cross section. However, we ignore
them since we believe that they do not change the following
conclusion. In the spin-isotropic case, we calculate
\begin{equation}
 \chi^{+-}(\bm{q},i\omega_n)=\int^\beta_0\mathrm{d}\tau
  \mathrm{e}^{i\omega_n\tau}
  \langle\hat{S}^+_{-\bm{q}}(\tau)\hat{S}^-_{\bm{q}}\rangle
\end{equation}
where $\hat{S}^+_{\bm{q}}$ is defined as 
\begin{equation}
\hat{S}^+_{\bm{q}}=\frac{1}{N}\sum_{\bm{k}}\sum_{a}
 c^{\dagger}_{\bm{q}+\bm{k}a\uparrow}c_{\bm{k}a\downarrow},
\end{equation}
and $a$ denotes the orbital indices. 
Then, $S(\bm{q},\omega)$ is obtained using the fluctuation-dissipation
theorem as
\begin{equation}
 S(\bm{q},\omega)=\frac{\mathrm{Im}\chi^{+-}(\bm{q},\omega)}
  {1-\mathrm{e}^{-\beta\omega}}.
\end{equation}

% overall view
The main panel of Fig.~\ref{fig1} shows the overall view of
$S(\bm{q},\omega)$ calculated
at $n=6.0$ and temperature $T=0.02$eV along the high symmetry lines of
the Brillouin zone shown in Fig.~\ref{fig1} (a). Note that the energy
range is beyond the experimental reach. We can
see two prominent features in this
figure: one is the peak at $\bm{q}\sim(\pi,\pi)$ and $\omega\sim 1$eV
and another is the peak at $\bm{q}\sim(\pi,0)$ and $\omega\sim
0$eV. Considering the DOS (shown in the inset of the main panel of
Fig.~\ref{fig1}) and dispersion relations, we can see that the
former corresponds to a transition between two van Hove
singularities above and bellow the Fermi energy. The latter peak comes
from the nesting of the Fermi surface. Remember that the wave vector
$(\pi,0)$ in the present model corresponds to $(\pi,\pi)$ in the
original Brillouin zone (Fig.~\ref{fig1}(b)).

\begin{figure}[t]
 \begin{center}
  \includegraphics[width=180pt]{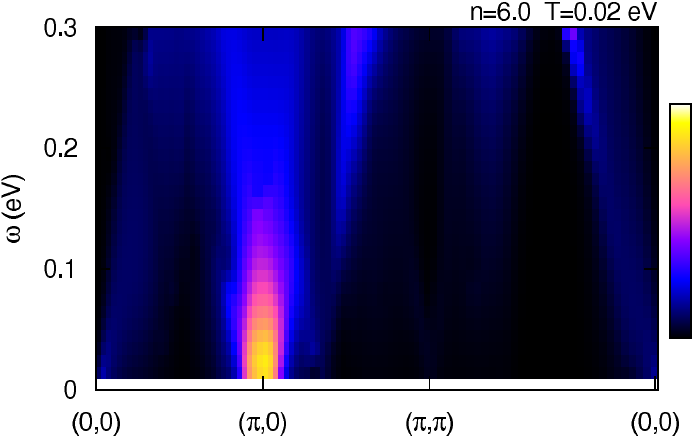}\\
  \includegraphics[width=180pt]{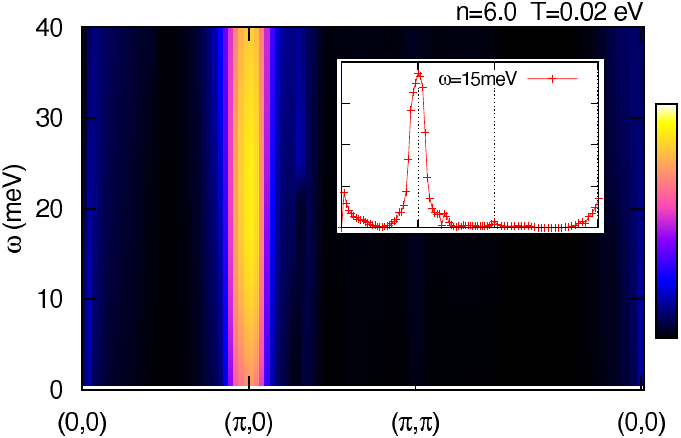}
  \caption{(Color online) Low energy close-up of $S(\bm{q},\omega)$ at
  temperature
  $T=0.02$eV (upper panel: up to 300meV, lower panel: up to
  40meV). Inset: $S(\bm{q},\omega')$ with $\omega'=15$meV.}
   \label{fig2}
 \end{center}
\end{figure}

% low energy part
The low energy part of $S(\bm{q},\omega)$ is shown in Fig.~\ref{fig2} to
compare with experimental results. We can see that the intensity
distributions above $(\pi,0)$ are almost vertical in this energy range,
which is similar to the experimentally observed $S(\bm{q},\omega)$. 
Quite often, the column-like response at $(\pi,0)$ is interpreted by
localized spin model. However, the present results indicates that the
localized spin model is not necessary to explain the spin dynamics of
this system. In principle, $S(\bm{q},\omega)$ in an itinerant model has
low-lying excitations for $0\leq |\bm{q}| \leq 2k_F$. However, in the
present model for iron-pnictides, there are disconnected small Fermi
surfaces around $(0,0)$ and around $(\pi,0)$. In addition to this,
since $2k_F$ is small, $S(\bm{q},\omega)$ has a rather narrow
column-like response at $(\pi,0)$. Furthermore, the moderate strength of
Coulomb interaction makes
the response more sharper in RPA treatment.
Actually, calculated peak width is estimated to be about 0.125 of the
distance between $(0,0)$
and $(\pi,0)$ in the extended Brillouin zone (In Fig.~\ref{fig2},
$S(\bm{q},\omega')$ with $\omega'=15$meV is plotted as inset). This width is
in fairly good agreement with experimentally observed width\cite{Matan:2008}. 
Although the most reported data (not all\cite{Ewings:2008}) are lower
than about 25meV at present, we make a plot of the same data up to 300meV in
the upper panel of Fig.~\ref{fig2}. We can see that the peak intensity
vanishes with increasing frequency in this energy range and also
that the V-shape peak structure becomes visible above 0.1eV.
Here, in comparing the calculated result with the experiments, we must
note that the band renormalization effect is neglected within our RPA
treatment. Thus, the actual energy range will be lower than that
indicated in Fig.~\ref{fig2}, since the dispersion relation is squished
by the band renormalization effect. Since the obtained peak intensity is
so prominent at $(\pi,0)$, the result does not change if we take into
account polycrystalline nature.

\begin{figure}[t]
 \begin{center}
  \includegraphics[width=140pt]{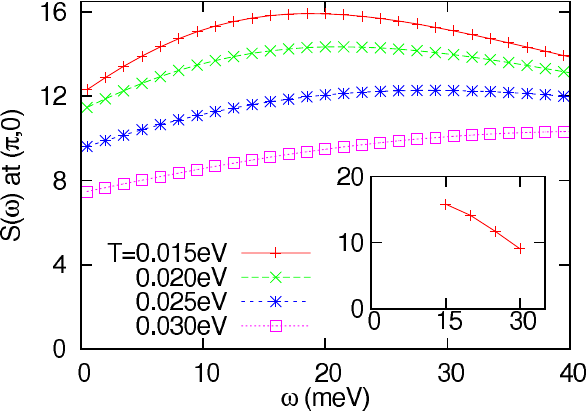}\\
  \includegraphics[width=140pt]{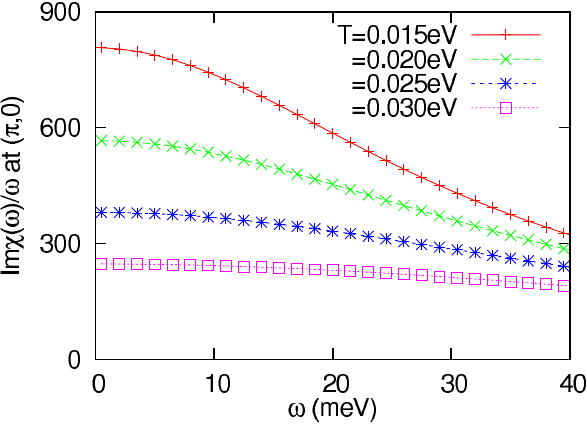}
  \caption{(Color online) Frequency dependence of $S(\bm{Q},\omega)$
  (upper panel) and
  $\mathrm{Im}\chi^{+-}(\bm{Q},\omega)/\omega$ (lower panel) at
  $\bm{Q}=(\pi,0)$ for several temperatures. Inset of upper panel: 
  Temperature dependence of $S(\bm{Q},\omega')$ at $\bm{Q}=(\pi,0)$ and
  $\omega'=15$meV. Note that horizontal axis of inset is temperature
  (meV).}
  \label{fig3}
 \end{center}
\end{figure}

% temperature and frequency dependence
In order to obtain a clearer view, the $\omega$-dependence of
$S(\bm{Q},\omega)$ at $\bm{Q}=(\pi,0)$, is shown in the upper panel of
Fig.~\ref{fig3}. 
We can see that
$S(\bm{Q},\omega)$ grows gradually with decreasing temperature. To see
this behavior, the temperature dependence of the peak intensity,
i.e. $S(\bm{Q},\omega')$ at $\omega'=15$meV is plotted in the inset of
Fig.~\ref{fig3}. This result is similar to the experimentally
observed temperature dependence of the peak
intensity above the ordering temperature\cite{Ishikado:2008}. We also
plot $\mathrm{Im}\chi^{+-}(\omega,\bm{Q})/\omega$ in the lower panel of
Fig.~\ref{fig3} to compare it with the other experiment by McQueeney
{\it et al.}\cite{McQueeney:2008}. This behavior
is also consistent with the experimentally obtained result above
ordering temperature\cite{McQueeney:2008}, considering the
renormalization effect on energy scale. These results
on temperature and frequency dependence is usual in RPA theory of
magnetism.
Here, we make a technical remark. Shown frequency dependence is obtained
by Pad\'{e} approximation, but we also calculate the same quantity in
another method where analytic continuation is not needed. Two results
are basically same, but Pad\'{e} approximation gives a little smoother
frequency dependence. 

% doping
Here, we briefly explain how above picture would change with
doping. With electron doping, the peak position around $\omega=0$ shifts
to the incommensurate position, where the peak of the static
susceptibility is found\cite{Kuroki:2008}. Namely, the main peak 
shifts in the direction from $(\pi,0)$ to $(\pi,\pi)$. As expected, the
scattering intensity of the low energy excitations quickly weakens by
doping since the doping breaks the nesting condition and brings the system
away from the magnetic instability.

% discussion
All the above results show that the itinerant model gives natural
explanation for the neutron scattering experiments at least in the
state without long-range order, although the sharp peak near $(\pi,0)$
has been often analyzed by local spin models. 
Here, we want to make some comments on the ordered state. First is about
how $S(\bm{q},\omega)$ looks like in the ordered state. If we consider
only Stoner type excitations as in the previous treatment, and if there
appears an SDW gap in the ordered state, low energy electronic
excitation is removed and $S(\bm{q},\omega)$ intensity drops for
$\omega$ in the SDW gap. This scenario may explain the observed
temperature dependence of peak intensity of $S(\bm{q},\omega)$
shown in Ishikado {\it et al.}\cite{Ishikado:2008}. However, in the
ordered state, we must
consider collective excitations, i.e., spin wave, as well. Then, it is
not so simple to reveal the actual shape of $S(\bm{q},\omega)$ in the
ordered phase. Second is about the order of
the phase transition, i.e., second order or first order. In our
treatment, or in the standard RPA approximation, transition becomes second
order. However, in the actual compounds, lattice deformation is
accompanied
and this may bring the transition to first order. Third, the
consideration on the three dimensionality is required to discuss the
ordered phase precisely. However, as shown by Matan {\it et
al.}\cite{Matan:2008}, the spin dynamics is highly two dimensional above
the ordering temperature, which justifies the present treatment.

\begin{figure}[t]
 \begin{center}
  \includegraphics[width=180pt]{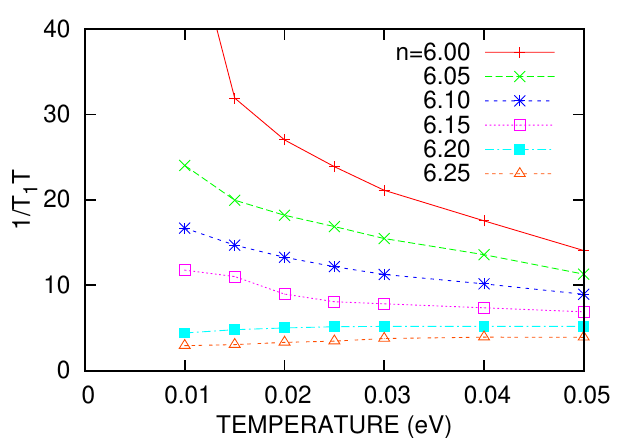}
  \caption{(Color online) Temperature dependence of $1/T_1T$ for several
  doping levels.}
  \label{fig4}
 \end{center}
\end{figure}

% NMR
Finally, we calculate the NMR result of $1/T_1T$ on the same footing,
i.e., itinerant model and RPA approximation. Figure~\ref{fig4} shows the
temperature dependence of $1/T_1T$ obtained in the present treatment.
Actually, Ikeda has calculated the normal state $1/T_1T$ in the FLEX
approximation, in which the self energy correction is also taken into
account\cite{Ikeda:2008}, and the arguments similar to our treatment are
presented in Graser {\it et al.}\cite{Graser:2008}. Thus, the present
calculation for $1/T_1T$ is just for checking the parameter choice.
The obtained result in Fig.~\ref{fig4} basically captures the
experimentally observed properties\cite{Nakai:2008b,Ning:2008a}. In the
lower doping side, $1/T_1T$ shows
enhancement with decreasing temperature due to the Fermi surface
nesting. On the other hand, in the higher doping side, $1/T_1T$ shows slight
decrease, which comes from the disappearance of the Fermi surface
nesting as well as the DOS structure of this model. However, this result
is not enough to explain the experiments quantitatively. One is about
the doping dependence. In experiments, the behavior of $1/T_1T$ changes
from increasing to decreasing more rapidly with doping. Another is about
the behavior in the higher doping side. Observed decrease is
more prominent than the present calculation, and shows one order drop
from room temperature to superconducting transition
temperature. Although the FLEX approximation\cite{Ikeda:2008} gives
a better result on this issue, the effects beyond the rigid band shift
may be important.

%summary
In summary, it is shown that the itinerant model treated within RPA
approximation can consistently explain the inelastic neutron scattering
results such
as peak width or temperature dependence of the peak intensity. 
The NMR results can be also roughly reproduced on the same footing, but
quantitatively, it is not satisfactory yet. Although our conclusion is that
the itinerant model reproduces normal state spin dynamics well (at least
qualitatively), our calculation does not exclude the possibility of the 
localized-spin model. Further works will be necessary for determining
whether the strong correlation and/or Mott insulating behavior is
important in this iron-pnictide superconductors. 
%acknowledge

\section*{Acknowledgment}
We thank T.~J. Sato for stimulating discussions.

\input{sqw.bbl}

\end{document}

%% file: sqw.bbl
%%%%%%%%Bibiography Style File for JPSJ %%%%%%%%%%%%%
% Released on November 15, 1996: Version 1.00       %
% Copyright (C) 1996 by Shinsaku Fujita,            %
%                             all rights reserved.  %
%%%%%%%%%%Bibliography%%%%%%%%%%%%%%%%%%%%%%%%%%%%%%%